\documentclass[11pt, times]{article}

\usepackage{geometry}                
\usepackage{amssymb}
\usepackage{placeins}
\usepackage[numbers, sort&compress]{natbib}
\usepackage{varioref}	
\usepackage[colorlinks, citecolor=blue]{hyperref} 	
\usepackage{hypcap} 	
\usepackage{gensymb} 	
\usepackage{textcomp}	
\usepackage{enumerate}	
\usepackage[breakable]{tcolorbox} 

\title{Using an added liquid to suppress drying defects in hard particle coatings}

\author{Steffen B. Fischer$^\mathrm{a,b}$, Erin Koos$^\mathrm{a,\ast}$ }

\date{\small
    $^\mathrm{a}$KU Leuven, Soft Matter, Rheology and Technology, Department of Chemical Engineering, Celestijnenlaan 200f, 3001 Leuven, Belgium \\
    $^\mathrm{b}$Karlsruhe Institute of Technology, Institute for Mechanical Process Engineering and Mechanics, Karlsruhe, Germany 
    ~\\
    ~\\
    $^{\ast}$ e-mail: erin.koos@kuleuven.be \\
}

\begin{document}
\maketitle

\begin{abstract}
	\centerline{\includegraphics[width=\linewidth]{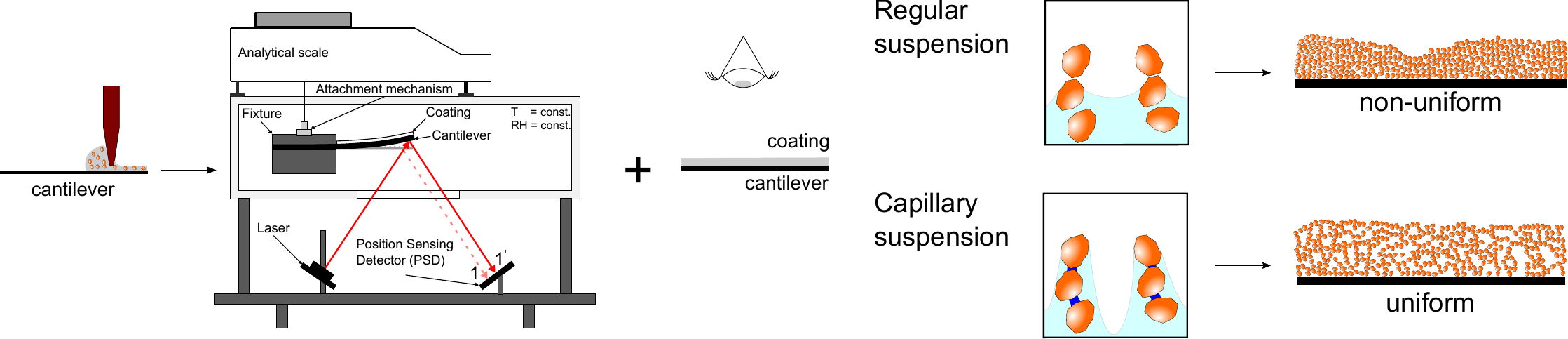}}
	\label{fig:graphical-abstractuniform-drying}
\emph{Hypothesis}:
Lateral accumulation and film defects during drying of hard particle coatings is a common problem, typically solved using polymeric additives and surface active ingredients, which require further processing of the dried film. Capillary suspensions with their tunable physical properties, devoid of polymers, offer new pathways in producing uniform and defect free particulate coatings. 

\indent\emph{Experiments}:
We investigated the effect of small amounts of secondary liquid on the coating's drying behavior. Stress build-up and weight loss in a temperature and humidity controlled drying chamber were simultaneously measured. Changes in the coating's reflectance and height profile over time were related with the weight loss and stress curve. 

\indent\emph{Findings}:
Capillary suspensions dry uniformly without defects. Lateral drying is inhibited by the high yield stress, causing the coating to shrink to an even height. The bridges between particles prevent air invasion and extend the constant drying period. The liquid in the lower layers is transported to the interface via corner flow within surface pores, leading to a partially dry layer near the substrate while the pores above are still saturated.  Using capillary suspensions for hard particle coatings results in more uniform, defect free films with better printing characteristics, rendering high additive content obsolete. 

\indent\emph{Keywords}: Drying; particle coatings; capillary suspension; stress measurement; lateral drying; corner flow; yield stress; shrinkage; uniform film 

\end{abstract}

\section{Introduction}
\label{Intro}

Particle coatings are widely used in various applications, such as protective paints and varnishes \cite{Chambers.2006}, production of thin film batteries \cite{Hu.2010}, printing RFID tags for the internet of things (IoT) \cite{Leenen.2009, Kamyshny.2014}, and printing solar cells at low cost \cite{Habas.2010}. In all these applications, correct functionality is indispensable. However, during drying of such coatings shrinkage occurs, which can cause cracks and other defects that can destroy their function. This can, for instance, allow mold growth in painted woodwork or rust in metals when they should be fully covered with protective paints and varnishes, or a loss of conductivity in the case of printed conductive pastes. This type of defect can also be observed in nature, for instance in the cracking of mud and silt \cite{Goehring.2010}. These problems are not just limited to cracks. An inhomogeneous dry film thickness in the calendaring step during Li-ion anode production can cause stresses leading to fracture, or to an undesired porosity distribution in the cross section of the electrode \cite{Schmitt.2014, Bitsch.2014}. 

Drying of soft and hard particle suspensions typically follow the same basic stages of drying \cite{Scherer.1990b}. Before their application, the functional particles are suspended in a continuous phase. Initially, in the constant rate period, the drying film behaves as if there were no particles present. After establishing the equilibrium wet bulb temperature, the drying rate stays constant as long as the surface of the film is covered with liquid. Within the coating, however, changes in the microstructure caused by transport mechanisms can have an influence on the coating properties. \citet{Cardinal.2010} mapped out three different regimes for colloidal suspensions: the sedimentation regime, in which particles settle in the liquid; the evaporation dominated regime where particles accumulate at the liquid-air interface, forming a consolidated layer and drying `top-down' until a maximum packing is reached; and the diffusion controlled regime, which opposes the previously mentioned effects and forms a uniform coating. 
After the constant rate period, the next drying stage occurs when the film has compacted. Further evaporation forms pinned menisci between the particles on the interface. These menisci deepen and increase the capillary pressure in the pores. The pressure, $ p_{c} $, can be described by the Young-Laplace equation: 
\begin{equation}
\label{eq:p_cap}
p_c= \frac{2\gamma_\mathrm{lv} \cos{\theta}}{r}
\end{equation}
where the capillary pressure depends on the liquid-vapor interfacial tension, $ \gamma_\mathrm{lv} $, the contact angle $ \theta $, and the radius $ r $ of a capillary tube inscribing the neck between particles. 

As drying continues and the liquid-vapor interface, also called drying front, recedes into the porous body, the drying rate decreases due to added vapor diffusion resistance from the pores to the surface. In the initial drying stage, the capillary pressure contracts its surroundings and the film shrinks. Later, when the film is fully compacted the capillary pressure induces stresses. 
If the top layers shrink first, there is a strain mismatch with the layer adhering to the substrate, which renders the bottom layer incapable of shrinking. This stress mismatch during drying can lead to cracking \cite{Lei.2001, Lei.2002}. Cracking during drying can be prevented by adding binders that take up stresses, i.e.~sterically oppose the capillary pressure contracting the particles, resulting in a strong dry film. Another way of lowering the stresses in particle coatings is by decreasing the capillary pressure \cite{BohnleinMau.1992} e.g., by adding surfactants to lower the water-air interfacial tension \cite{Chiu.1993b, Bauer.2009}, according to \autoref{eq:p_cap}. This approach works well as long as the additives can remain in the film and no subsequent treatment of the coating is needed, as is the case in paints. 
Surfactants and binders, however, can be problematic if the coating obtained after drying is merely an intermediate step, such as a ceramic green body, where a subsequent pyrolysis step is needed before sintering. The same is true for printed electronics where neck formation between particles aids in the formation of an electrically conductive percolated network. The burn out can cause two major problems. First, if the pyrolysis conditions are not properly identified, organic residues may remain in the film, which can reduce the efficacy of the coating \cite{Tsai.1991}. Second, since the polymers occupy space between the particles, non-uniform burn out can cause shrinkage and therefore the introduction of additional stresses and possible cracks \cite{Fu.2015b}. 

Another phenomenon occurring during drying of particle laden droplets and coatings is lateral drying. Deegan et al.~\cite{Deegan.1997} have identified capillary flow as the root cause for the coffee-ring effect in which suspended coffee grounds accumulate at the edges of a drying droplet. The underlying mechanism is the pinning of the air-liquid-substrate contact line, which prevents a contraction of the interface towards the center. In order to balance evaporation losses, a fluid flux towards the pinned contact line occurs. This capillary flow has also been observed in film drying and manifests as lateral drying of coatings as observed by many researchers \cite{Chiu.1993b,Guo.1999,Holmes.2008,Ma.2005,Routh.1998}. These capillary flows can lead to cracking as well as drying defects such as dimples or particle depleted trenches \cite{Holmes.2006}.

As previously mentioned, cracking can be diminished by the addition of soft polymers or surfactants, but this may cause problems or require additional processing steps for some applications . To reduce the amount of additives, Jin et al.~\cite{Jin.2013} added a surfactant stabilized emulsion to a colloidal suspension. The effect was an increase in the shear moduli with a corresponding dramatic change in cracking. Another concept for suspension formulation uses capillary suspensions \cite{Koos.2011,Koos.2014,Koos.2014b}, where a small amount of a liquid immiscible with the bulk fluid was added to a suspension, which resulted in a dramatic change of its yield stress and elastic properties. In our previous work \cite{Schneider.2017}, we observed a remarkable decrease in crack formation for capillary suspensions even in the absence of surfactants. 
This paper investigates this phenomenon using a stress measurement apparatus to examine the stresses that are responsible for cracking. We compare the obtained stress curve with the visually observed drying features and put forward a hypothesis for the observed changes in the drying behavior of the capillary suspension in comparison to the regular suspension without added secondary liquid.

\section{Materials and Methods}
\subsection{Sample preparation}

Two different suspensions were formulated, differing only in the addition of a small amount of water. Alumina particles ($\alpha$-Al$_2$O$_3$, CT3000SG, Almatis GmbH, Germany) with an average particle size of d$ _{50,3}=0.5 $~{\micro}m according to the supplier, were chosen for the solid phase. For the continuous phase, 1-heptanol ($>$99\%, Alfa Aesar) was used. Particles were dispersed in multiple steps (at least twice for two minutes at 3500 rpm) with a Speedmixer DAC 150.1~FV (Hauschild \& Co. KG, Germany) to obtain smooth samples with particle volume fractions of $\phi_\mathrm{Al_2O_3}=0.2$. To form capillary suspensions, ultra-pure water (atrium 611 DI, Sartorius AG, Germany) was added to achieve a water volume fraction of $\phi_\mathrm{sec}=0.025$, while maintaining a solid content of $\phi_\mathrm{Al_2O_3}=0.2$. The specimens were mixed at 3500~rpm in increments of two minutes, to prevent the samples from excessively heating. The mixer is designed in such a way that degassing after mixing is not necessary. The samples were stored in the polypropylene Speedmixer cups wrapped in parafilm until use. Films were coated onto stainless steel cantilevers, as detailed in the Supplementary Information. 

For some experiments the samples were dyed in order to trace concentration gradients. Before mixing, a stock solution of 1-heptanol was prepared by adding trace amounts (266 ppmw) of oil red EGN (Sigma-Aldrich, Germany). It is highly soluble in oils and appears red in the form of powder as well as in solution, while practically immiscible in water. The secondary phase, water, was dyed with a stock solution (50 ppmw) of fluorescein, disodium salt (ACROS Organics, Belgium), which is UV active. It is highly soluble in water and immiscible in the oil. The dye appears red when dry, whereas it fluoresces in solution under UV light to appear yellow-green. For these experiments, a handheld UV lamp was added to the experimental setup. 

\subsection{Stress measurement and drying chamber}

The stresses during drying are measured using the cantilever deflection method where bending of the substrate due to shrinkage of a well-adhered film is recorded with a laser. The drying rate is more difficult to obtain. Usually, a second cantilever is simultaneously coated and inserted in the drying chamber in order to link stress development and drying rate \cite{Lei.2001}. However, small differences of the coatings and drying conditions may lead to inaccuracies in attributing observed changes. To overcome these drawbacks, we designed a setup such that it is capable to simultaneously measure the stresses and drying rate of the same sample. The fixture with the sample coated cantilever (\autoref{fig:method}a) is placed in a temperature and humidity controlled chamber (\autoref{fig:method}b). The fixture is suspended from a scale while measuring the cantilever deflection with a laser. Drying was carried out in dry air at 40~{\degree}C with approximately 1\% relative humidity (RH).

For this work, we placed a USB camera, lit with LEDs through a hole in the top of the chamber, below the scale. This camera imaged the front part of the cantilever, away from the clamp. Video recording was started simultaneously with the stress measurement to ensure comparability. For experiments with undyed samples, the LED light of the camera was turned on to illuminate the specimen. For the dyed samples, we used indirect lighting through the chamber door, which is made of PMMA. Due to the restricted space available for the camera and its lack of wide angle capabilities, we only recorded the first 27~mm of the total coating length of 40~mm. The snapshots were processed in ImageJ \cite{Schneider.2012}. The image intensity was adjusted using an adjacent area that was not subject to drying in order to account for changes in light or camera shutter adjustments. Subsequently, the obtained gray values were contrast enhanced with a constant color palette for each measurement, referred to as false color in the figures below. 

\subsection{Film shrinkage}
For the film shrinkage measurements, we used the drying setup as described in \citet{Jaiser.2016}. The drying chamber constitutes a temperature controlled movable plate, which was set to 30~{\degree}C or 40~{\degree}C. Relative humidity could not be controlled in this experiment. A laser profiler (LJ-G015, Keyence, Japan) with a line width of 7~mm covering the whole scanning width of the cantilever (6~mm) is mounted to the chamber. The impinging air flow was shut off in order to achieve comparable conditions as in the drying chamber used for stress measurements. After coating the cantilevers, we placed them on the heated plate and measured the profile thickness in intervals of ten and five minutes for the lower and higher temperatures, respectively. It is noted that the drying conditions from a heated plate differ somewhat from a drying chamber. While drying times may be different, the qualitative change of shrinkage should remain similar.

\subsection{Surface profilometry}
Four cantilevers were coated with a regular suspension as well as a capillary suspension with $\phi_\mathrm{sec}=0.025$ according to the previously mentioned method, and subsequently dried in the stress measurement chamber at temperatures of 30~{\degree}C and 40~{\degree}C at 1\%~RH. Afterwards, their film thickness profile was measured with a DektakXT stylus surface profilometer (Bruker, USA). A map scan was performed with a resolution of 100~{\micro}m per trace  spanning 5000~{\micro}m. The examined cantilever area was chosen such that it reflected special features of the dry film.  

\section{Results}
\subsection{Comparison of stress curve features with drying images of a coating} 

\begin{figure}[htp]
	\capstart
	\centering
	\includegraphics[width=1\columnwidth]{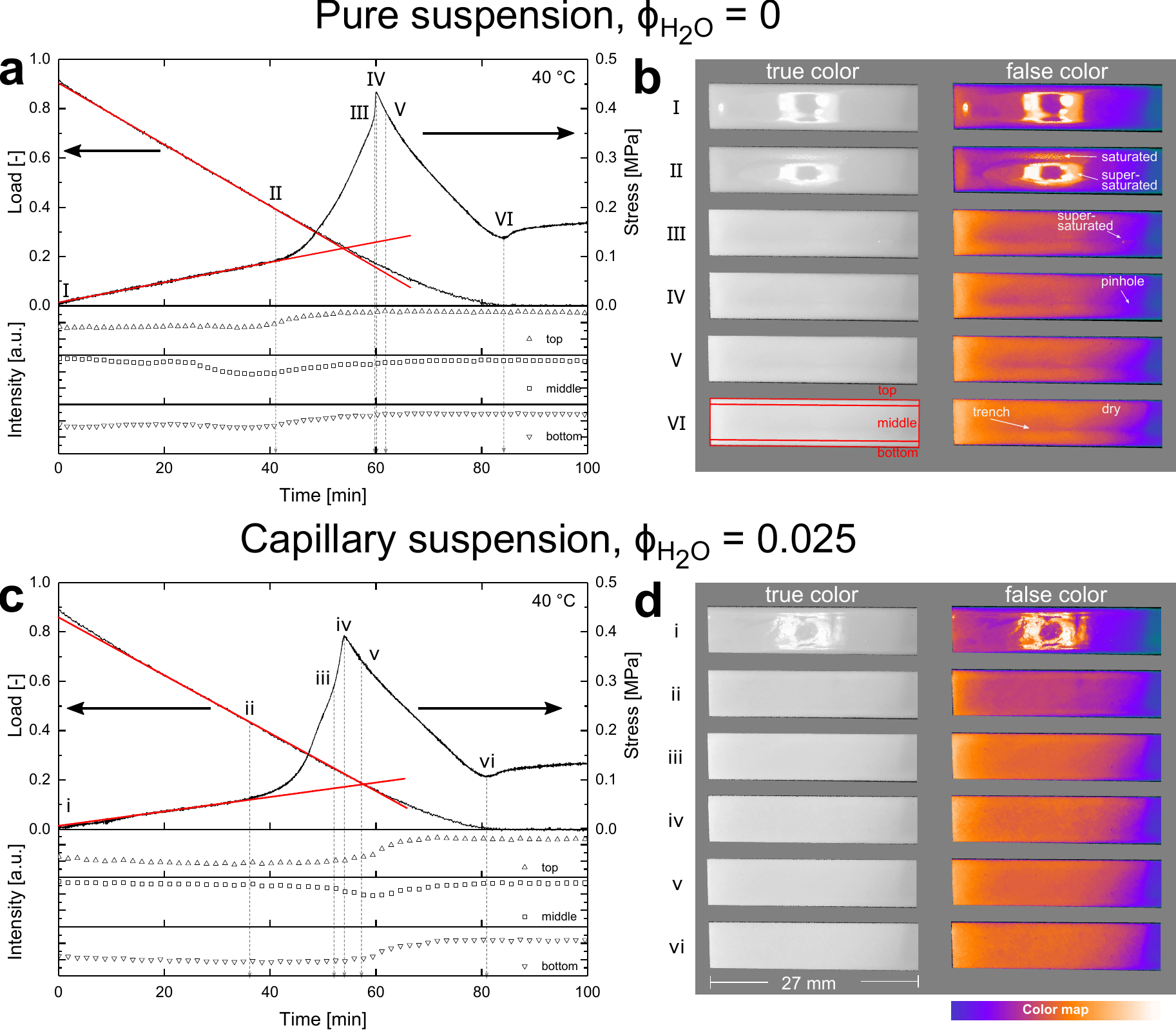}
	\caption{ Results for (\textbf{a},\textbf{b}) pure suspension without added water  and (\textbf{c},\textbf{d}) capillary suspension with $\phi_\mathrm{sec}=0.025$. The graphs in (\textbf{a}) and (\textbf{c}) show the load (top, left axis) and film stress (top, right axis). The lower panels describe the captured intensity change in three segments of the investigated coating (marked in b.VI). The top and bottom segments are narrower to capture lateral drying, whereas the middle segment is an average value of the remaining observable coating. The red lines mark the tangents in the constant rate period. The photographs in (\textbf{b}) and (\textbf{d}) show the cantilevers at different drying stages, marked in the graphs with Roman numerals. The films are directly lit with the camera's LED light, which is visible in the form of a circular ring of light that is reflected by a region of supersaturation as shown in b.II. The left column shows the true color image. In the right column, the true color image is converted to gray scale values and subsequently contrast enhanced with white showing bright areas and the dark areas in blue as depicted in the color map below (\textbf{d}). Both experiments were conducted at 40~{\degree}C and the scale is the same for all images. Videos of the false color images shown in (b) and (c) are available as Video 1 and Video 2, respectively.}
	\label{fig:undyed}	
\end{figure}

The stress profile, image intensity and photographs of the coatings at different drying times are shown in \autoref{fig:undyed} 
for both the regular suspension in the upper row (a,b) and for the capillary suspension with an added water content of 2.5 vol\%  in the lower row (c,d). \autoref{fig:undyed}a and \autoref{fig:undyed}c show the film stress profiles, as measured during the drying process at 40~{\degree}C. The stress values are shown on the top panel right y-axis, and change in mass, expressed as the load on the left y-axis. The time-variant load of the film is defined as:
\begin{equation}
\label{eq:load}
\mathrm{load}=\frac{m(t)-m_\mathrm{final}}{m_\mathrm{final}}=\frac{m_\mathrm{liquid}(t)}{m_\mathrm{dry}}
\end{equation}
and can therefore, be greater than unity. Since the load depends on the gravimetrically determined difference in evaporated liquid, the load can only be compared if the attained films have similar residual fluid content in the film, as is the case for the films investigated in this work. Otherwise, a larger amount of residual fluid is attributed to the mass of the dry film. We further restricted these experiments to average dry film thicknesses of 75~$\pm$~5~{\micro}m to ensure the comparison of films without cracks and interference of other mechanisms. Additionally, we avoid bias introduced through stress dependence on film thickness as observed by \citet{Tirumkudulu.2004}.
We identified six different points in time according to the following criteria, labeled with Roman numerals (uppercase for the pure suspension and lowercase for the capillary suspension). 
\begin{enumerate}[I:]
	\item the start of the measurement;
	\item the point where the stress increases faster than the initially linear slope (red tangent);
	\item the rapid increase in stress preceding the peak stress;
	\item the peak stress of the measurement;
	\item the rapid decrease in stress following the peak stress;
	\item a trough as the stress approaches the residual value.
\end{enumerate}

\autoref{fig:undyed}b and \autoref{fig:undyed}d show photographs of the cantilevers at these key points. The images on the right have been recolored from blue to bright yellow (colormap below \autoref{fig:undyed}d) to better show gradients and differences between images. These images are also available as Video 1 (\autoref{fig:undyed}b) and Video 2 (\autoref{fig:undyed}d). The intensity span for the color map was maintained within and between samples. The change in color is also depicted on the lower panels of \autoref{fig:undyed}a and \autoref{fig:dyephotos}c, which show the intensity values as an average value for narrow strips on the cantilever's edges as well as the middle of the cantilever (as marked in \autoref{fig:undyed}b.VI).  
The behavior of films during drying is shown in \autoref{fig:undyed}. The coating is illuminated using direct LED lighting, visible as a circular ring in the glossy liquid coating in \autoref{fig:undyed}b,d, which amplifies intensity changes caused by drying features. 

The stress curves for the pure suspension ($\mathrm{\phi_{sec} =0}$, \autoref{fig:undyed}a) exhibits the typically observed stress increase, as labeled on the right y-axis \cite{Chiu.1993b}. There is a gradual stress increase up to point II, which coincides with the constant rate drying period. At that point, the tangent, shown in red, starts deviating from the constant mass loss (left y-axis). The images of the cantilever show that the first clear change in intensity appears in the middle segment for the regular suspension after about 25 minutes when the average intensity curve drops until point II is reached. This decrease is linked to the lateral drying, which reduces the reflected area. This decay is not visible at the edges, likely due to the curvature of the film near the edges (non-reflective narrow top and bottom segments in image I). However, we observe an intensity increase in all three segments near point II, when the stress increase accelerates beyond the linear slope. At this point, we see an increase in the reflection when the saturated (compacted) edges start to reflect the light (along and across the cantilever) and are superimposed with the decreasing reflection of the supersaturated area predominantly visible in the middle section. This observation can clearly be seen in the color enhanced image II in the center of the top segment. After that, the total intensity asymptotically increases and no further conclusions can be drawn from an intensity change. The supersaturated area subsequently decreases in width and length along the cantilever. The interface between the supersaturated and saturated regions has a pointy shape as shown in \autoref{fig:undyed}b.II. As the supersaturated region retracts towards the center of the cantilever, a particle depleted area, i.e.~a trench \cite{Holmes.2006}, forms at its tip (as marked in \autoref{fig:undyed}b.VI). A dimple, or pinhole (\autoref{fig:undyed}b.IV), forms when the final droplet shaped supersaturated area disappears, as denoted in the contrast enhanced image \autoref{fig:undyed}b.III. The color changes towards yellow, indicating a dryer area, from the outside in as lateral drying continues. The trench and pinhole remain darker, making the contrast more visible.

The capillary suspension ($ \mathrm{\phi_{sec}=0.025} $) results are displayed in \autoref{fig:undyed}c,d. Comparing the reflection of both suspensions in the first true color image, the surface of the capillary suspension appears rougher (scattered reflection), possibly due to the absence of a liquid film on the surface. As drying proceeds, there is no clear change in intensity in all three segments of the capillary suspension until point ii is reached. While there is barely a change in intensity between points i and ii, the comparison of the images at the respective points reveals a difference. Even at point ii, which is reached earlier in the capillary suspension (36 min compared to 41 min in the pure suspension), the surface of the coating is no longer glossy and looks dry and devoid of a liquid surface film. However, the film is still in the constant rate period. Contrary to the regular suspension, the intensity of the middle segment decreases at point ii when the stress increase accelerates rather than increases. This continues until after point v, where the intensity in all three segments increase again. However, there is no clear link between this increase and any distinct points in the stress curves or images. After the peak, at point v, we reach the most obvious difference between the stress response for the two formulations. The constant rate period for capillary suspensions appears to be prolonged until point v, indicating the retraction of the liquid level into the film \emph{after} the consolidation point, which is reached at the peak stress (point iv) according to the common theory of drying \cite{Guo.1999}.

\subsection{Dyed samples}

\begin{figure}[htp]
	\capstart
	\centering
	\includegraphics[width=1\linewidth]{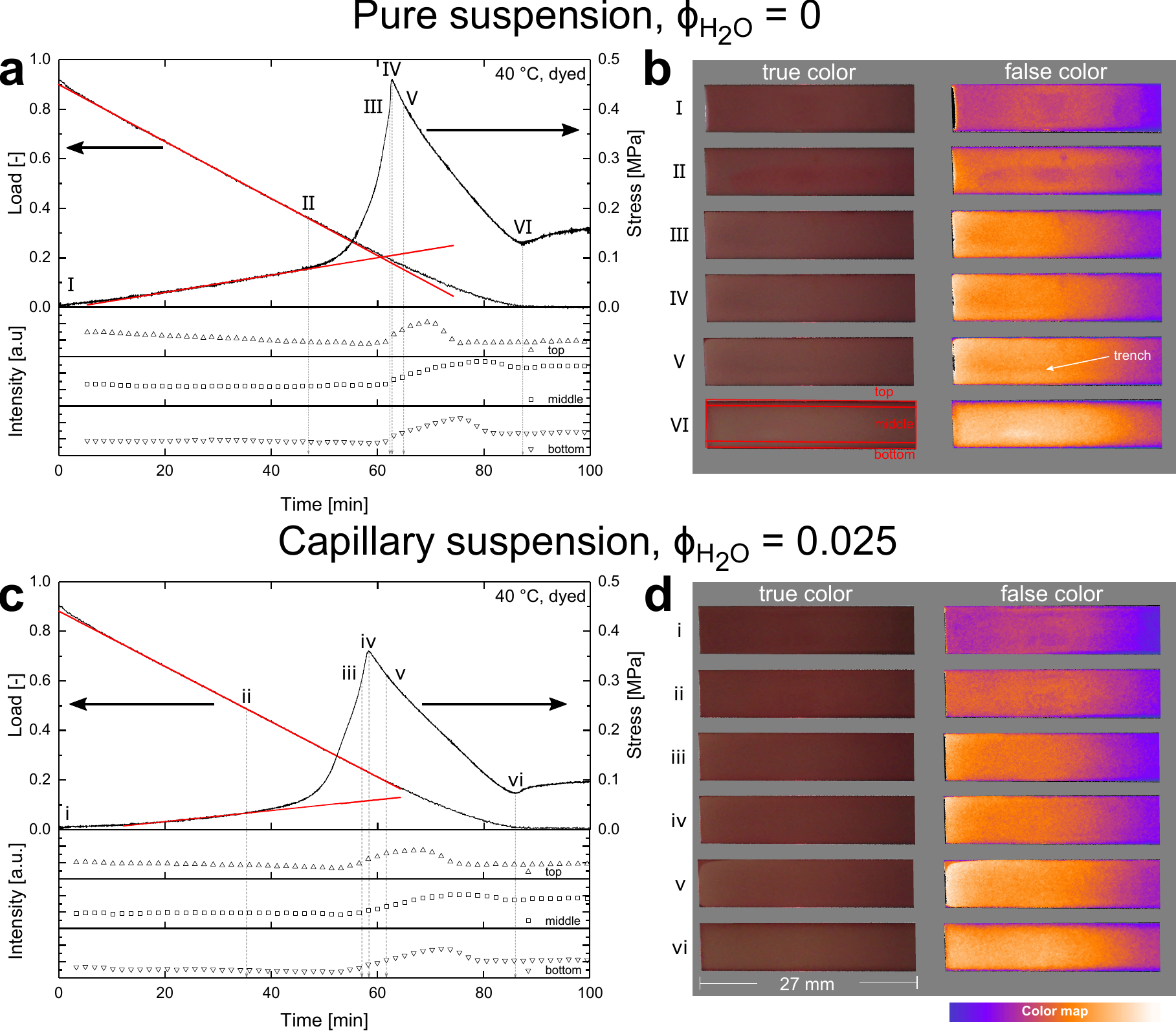}
	\caption{ Results for experiments with dyed bulk heptanol phase showing (\textbf{a},\textbf{b}) pure suspension without added water and (\textbf{c},\textbf{d}) capillary suspension with $\phi_\mathrm{sec}=0.025$. The graphs in (\textbf{a}) and (\textbf{c}) show the load (top, left axis) and film stress (top, right axis). The lower panels describe the captured intensity change in three segments (marked in b.VI) of the investigated coating. The top and bottom segments are narrow to capture lateral drying, whereas the middle segment is an average value of the remaining observable coating. The red lines mark the tangents in the constant rate period. The photographs in (\textbf{b}) and (\textbf{d}) show the cantilevers at different drying stages marked in the graphs with Roman numerals. The left column shows the true color image. In the right column, the true color image is converted to gray scale values and subsequently contrast enhanced with white showing bright areas and the dark areas in blue as depicted in the color map below (\textbf{d}). Both experiments were conducted at 40~{\degree}C and the scale is the same for all images.}
	\label{fig:dyephotos}
\end{figure}

To examine the mobility of the bulk liquid relative to the particles, experiments with dyed heptanol are shown in \autoref{fig:dyephotos}. 
The stress response of these measurements resembles the undyed samples shown in \autoref{fig:undyed}, albeit with a slight shift in the key times and maximum stress intensities. In the drying of the dyed regular suspension without added water (\autoref{fig:dyephotos}b), the image intensity values (bottom curves) show a clear decrease for the upper edge of the cantilever (top rectangle), which indicates lateral drying that is more pronounced from the upper edge. A decrease in intensity can be explained by lower light reflection from dry parts compared to a glossy wet surface. This change in intensity is indicated in \autoref{fig:dyephotos}b with a change in true color from darker red to a brighter red. The contrast enhanced picture depicts this behavior as a change from blue to more yellow. In II, a contrast difference marks the contour line of the supersaturated suspension. At point III, just before the final rapid stress increase (right y-axis), the intensity level in all three areas begin to rise. This intensity increase starts at approximately the same time across the cantilever denoting a reduction in the supersaturated area. This can be observed in terms of a color change towards more white on the true color or yellow on the false color images, respectively. Between point III and IV, which represent the final rapid stress rise, neither the averaged intensity values nor the images show any change in this short period of time. At point V, the initial stress release following the peak, the intensity increases but this change is gradual. In the images, however, a particle depleted area forming a trench down the center of the cantilever becomes visible. Additionally, the top left corner of the cantilever, which dried first, shows a darker red color. At point VI, there is an additional recognizable change of stress. Here, the stress reaches a local minimum, before increasing towards a residual stress value. Looking at the load, it is evident that the last of the heptanol begins to dry, further compacting the film. Moreover, this seems to be in accordance with a small minimum of the intensity in the middle segment of the cantilever. One explanation could be fluid transport from the inside of the film towards the surface, with subsequent precipitation of the dye changing the reflection intensity of the light. In the images, this change was too small to be observed in the center of the film. At the edges, on the other hand, it is clearly more red than in the center of the cantilever. Looking at the false color image in \autoref{fig:dyephotos}b.VI, this is clearly visible as narrow blue area at the edges, transitioning to yellow with an almost white patch in the lower left middle segment. 

Comparing the capillary suspension (\autoref{fig:dyephotos}c,d) with the sample without added liquid (\autoref{fig:dyephotos}a,b), we see particular differences. The stress increase (right y-axis) deviates from the constant rise (point ii) at an earlier time (35 min) for the capillary suspension than the regular suspension (47 min). At this time, the drying rate of the capillary suspension (left y-axis) is still constant, which is in contrast to the regular suspension where the stress and load curves both deviate at point II. There is also less change in the intensity of the top edge; the change of intensity between the beginning of the measurement and point ii appears to be more uniform. Between points ii and iii, the intensity increase, at approximately 55 minutes coincides with an inflection point in the stress curve. Here, the start of the intensity gain is not related to the final stress increase (point iii) near the peak but takes place about five minutes earlier. At point iii, we observe a moderate continuation of the stress increase towards the peak with a lower slope than for $\phi_\mathrm{sec}=0.025$. The contrast keeps gradually and uniformly changing albeit remaining darker near the edges. The capillary suspension coating does not show a sign of trench forming near the peak stress at point iv. However, drying appears to be lateral as well, observable from the darker top corner in the images at point v. The prolonged increase of the middle segment intensity supports this observation. At point vi, the middle segment seems to exhibit a small local intensity minimum in accordance with the local stress minimum. The final capillary suspension coating shows some dye accumulation at the edges but a more uniform dye distribution across the cantilever. The dyed heptanol, which acts as the bulk phase of the suspensions, seems to be transported to the edges of the cantilever, a sign of lateral drying. This mass transport appears to be reduced for capillary suspensions.
 
\begin{figure}[t!]
	\capstart
	\centering
	\includegraphics[width=1\linewidth]{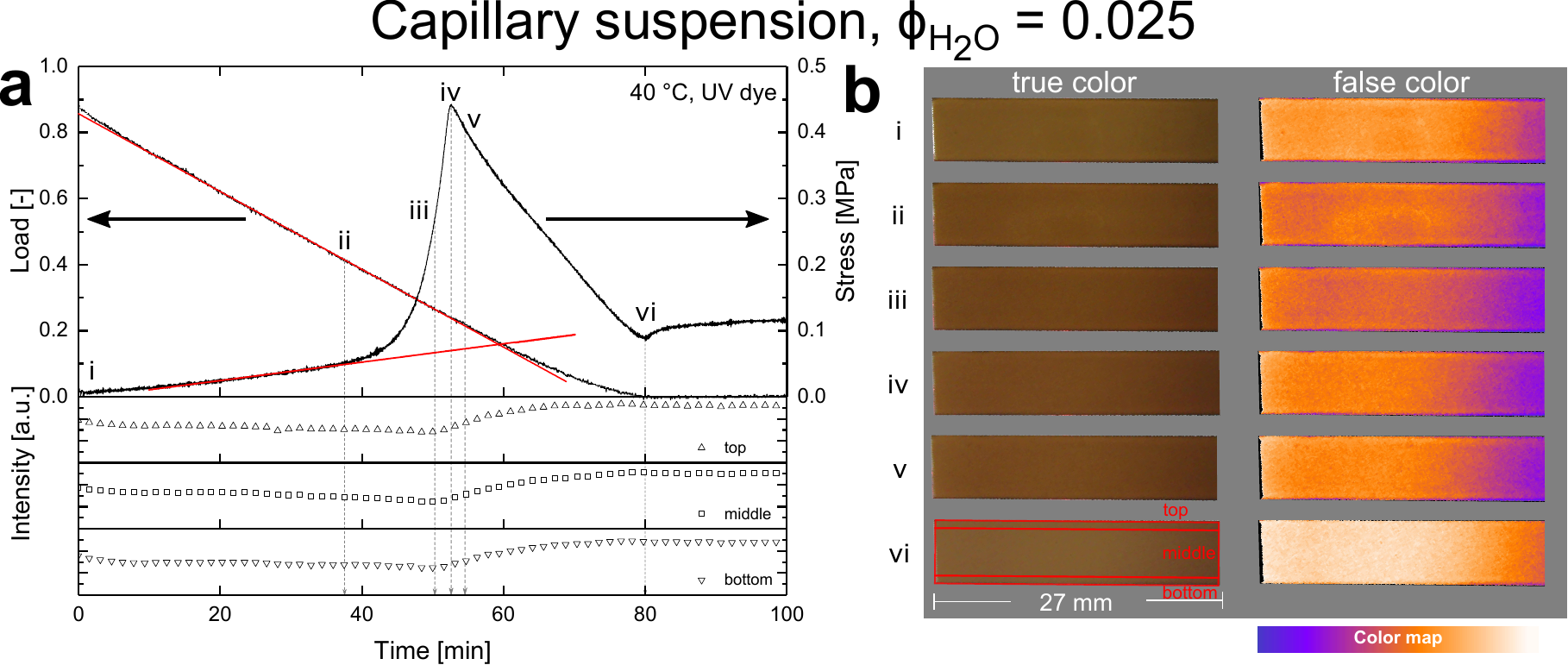}
	\caption{Results for experiments with dyed secondary water phase $\phi_\mathrm{sec}=0.025$. The graph in (\textbf{a}) shows the load (top, left axis) and film stress (top, right axis). The lower panels describe the captured intensity change in three segments (marked in b.vi) of the investigated coating. The top and bottom segments are narrow to capture lateral drying, whereas the middle segment is an average value of the remaining observable coating. The red lines mark the tangents in the constant rate period. The photographs in (\textbf{b}) show the cantilevers at different drying stages marked in the graphs with Roman numerals. The left column shows the true color image. In the right column, the true color image is converted to gray scale values and subsequently contrast enhanced with white showing bright areas and the dark areas in blue as depicted in the color map below (\textbf{b}). The experiment was conducted at 40~{\degree}C and the scale is the same for all images.}
	\label{fig:UV}
\end{figure}

In order to elucidate the mass transport behavior of the water liquid bridges, we dyed them yellow with a water soluble dye, as shown in \autoref{fig:UV}. 
In \autoref{fig:UV}a, the stress profile is very similar to the capillary suspension with dyed bulk phase in \autoref{fig:dyephotos}c. In this sample, point ii, marking the beginning of the stress rise, and the end of the constant rate period for the regular suspension, occur at nearly the same time as the previous capillary suspension sample, but the stress peaks at a slightly earlier time (53 min instead of 58 min) and has a higher value (0.45~MPa instead of 0.36~MPa). 
The intensity values shown in the lower panels of \autoref{fig:UV}a gradually decrease after the start of the measurement indicating a decreasing light reflection on the coating surface. The intensity begins to increase on the entire cantilever at, or shortly after, point iii when the rate of stress development subsides slightly. In contrast to the dyed bulk phase experiments, the dye intensity of the coating increases uniformly in all three segments, approaching their final values. This concurrent gain at the edges as well as in the middle part of the cantilever suggests a uniform drying where there is a uniform change over time across the cantilever. There is no visible accumulation of dye at the edges. This suggests that the capillary bridges, and thus the particles, are not transported towards the cantilever edges due to lateral drying.

Combined, the experiments in the drying chamber show key differences between the two formulations. The capillary suspension exhibits an extended constant rate period with the end of the linear regime occurring near the peak stress rather than near the initial stress rise as it does in the pure suspension. The images show that while both samples undergo lateral drying, this is reduced in the capillary suspension. Any lateral drying in the capillary suspension is also not associated with particle transport and particle depleted areas (trenches or pinholes) do not develop, unlike the pure suspension. Differences in the lateral drying are further explored by investigating the film profile, as shown below.

\subsection{Cross sectional profile change during drying}
\label{sec:profile}

\begin{figure}[t!]
	\capstart
	\centering
	\includegraphics[width=1\linewidth]{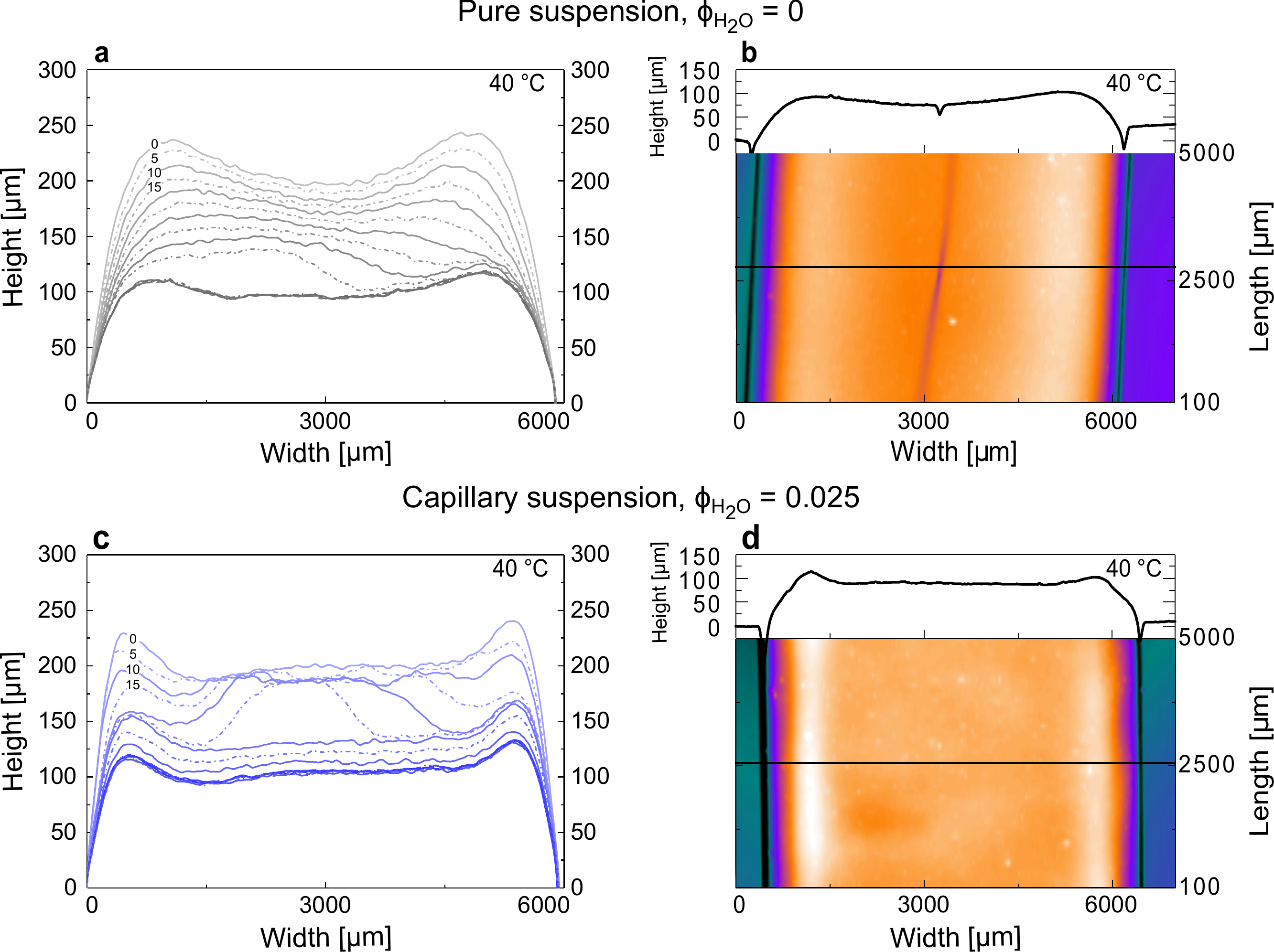}
	\caption{ Film profile measurements of (\textbf{a, b}) pure suspension and (\textbf{c, d}) capillary suspension. The laser profiles during drying, shown in (\textbf{a}) and (\textbf{c}) were taken every five minutes during contact drying at 40~{\degree}C and are displayed with solid and dashed lines in an alternating fashion. Increasing opacity of the curves indicate later stages of drying. The profiles in (\textbf{b}) and (\textbf{d}) capture a five millimeter surface section for films dried in the stress measurement chamber. The horizontal black line shows the cross sectional cut for the height profile in the upper panel.}
	\label{fig:shrinkage_40C}
\end{figure}

In the previous section, we observed a different drying behavior for capillary suspensions that suggests less prominent lateral drying. Additionally, light reflected and scattered on the coating surfaces implies a different topology. Therefore, we studied the change of a coating cross section for two suspensions over time, as shown in \autoref{fig:shrinkage_40C}a,c. 
Over the course of drying on a heated plate, the height profile was measured in intervals of five minutes. The profiles in the graphs are depicted with more color saturation, as well as alternating dashed lines for increasing time steps. After coating the regular suspension ($ \phi_\mathrm{sec}=0 $) on the cantilever, we observed a significantly elevated film at the outside of the cantilever (\autoref{fig:shrinkage_40C}a). These edge effects depend on the intrinsic properties of the coating, such as elasticity or interfacial tension, as well as extrinsic application properties, such as coating speed or technique \cite{Schmitt.2014}. As drying begins, the film shrinks vertically across the cantilever, but this shrinkage is more pronounced at the edges. After about 25 minutes, the edges are nearly the same height as the center, implying that the film undergoes lateral drying. Pinning of the contact line causes the contact angle between the film and cantilever to decrease while the film is shrinking. Afterwards, the lateral drying slows and the evaporation in the center again dominates. This evaporation is enhanced on the right hand side of the profile, perhaps due to unbalanced natural convection. After roughly 50 minutes, the final compacted film structure is reached. This point should correspond to the peak stress (point IV) in \autoref{fig:undyed}a and \autoref{fig:dyephotos}a with the differences in time arising from the different drying methods. The compacted profile has a more leveled surface shape compared to the wet state, but is still clearly lower in the center. This final, dense state is maintained during the remainder of the drying process. The profiles of a film dried at 30~{\degree}C, shown in the supplementary information \autoref{fig:shrinkage_30C}, exhibits similar behavior albeit without the asymmetric drying. The small peaks and valleys between profiles are measurement uncertainties of the device.

The profile of a dry film, which was dried in the environmental chamber, is shown in \autoref{fig:shrinkage_40C}b, where lower areas are blue and elevated regions are red. The black line denotes the location of the height profile. The suspension without added water shows a very similar dry profile as in \autoref{fig:shrinkage_40C}a with slight superelevations despite the different drying conditions. Here, we have highlighted a trench covering almost the entire length of the cantilever with a pinhole in it. The dimple, representing the deepest depression of the trench is shown in the black height profile in \autoref{fig:shrinkage_40C}b. Additionally, the dry film surface is uneven across the whole cantilever. When we dried the regular suspension at a lower temperature, see \autoref{fig:shrinkage_30C}, we did not observe a trench formation but the pinhole is still present. Comparing the dry film thickness for the regular suspension at 40~{\degree}C (\autoref{fig:shrinkage_40C}a) and at 30~{\degree}C (\autoref{fig:shrinkage_30C}) combined in (\autoref{fig:dry_film}c), we can see a more even surface in the center and steeper, more prominent superelevations towards the edges at the elevated temperature, even while starting from a very similar wet film profile. This observation is also visible in the samples dried in the drying chamber and measured with the stylus when comparing \autoref{fig:shrinkage_40C}b and \autoref{fig:shrinkage_30C}b. Drying of the regular suspension leads to particle depleted spots such as dimples or, at higher temperatures under more prominent lateral drying to trenches as already observed by others \cite{Holmes.2006}. 

The drying profiles of a capillary suspension are shown in \autoref{fig:shrinkage_40C}c,d. The initial wet coating profile is substantially different from the regular suspension. The coating edge angle is closer to 90° than for the regular suspension, resulting in a better shape accuracy \cite{Schneider.2016}. This shape accuracy is attributed to their high yield stress, shear thinning and excellent yield-recovery behavior \cite{Koos.2014b}. The capillary suspension does have sharp superelevations but the height profile between the high edges is flatter. The height of the superelevations and the difference between the center height are roughly equal to the suspension without added water. These superelevations are a fault of the coating process caused by the cantilever edges and the intrinsic properties, as discussed earlier. Upon the start of drying, lateral drying occurs where the superelevations at the edges dry first, shrinking while the center remains at the same approximate height. After 15 minutes, the edges of the film lower beyond the height of the center as the lateral drying front propagates inward. In contrast to the regular suspension, the supersaturated wet area in the center remains at the same thickness such that a hill is formed. This persists until the superimposed lateral drying rates meet and the highest point in the center starts to evaporate. After 30 minutes, the hill on the surface has fully evaporated and the center once again forms a flat profile. This time corresponds to the beginning of the stress rise (point ii in \autoref{fig:undyed}, \autoref{fig:dyephotos}, and \autoref{fig:UV}), but is not coincident with the end of the constant rate period. Afterwards, the surface maintains its shape, while continuing to decrease in total height. Once the surface center is flat, it barely changes its form and appears to settle as a uniform body until the final thickness is reached. The final profile is obtained after 45 minutes. As with the pure suspension, this point where the film is fully compact should correspond to the stress peak (point iv). The profile of the dry film, \autoref{fig:shrinkage_40C}d, shows the superelevations and an otherwise flat profile devoid of defects. Unlike the pure suspension, there is no dimple or trench. This is remarkably different from the regular suspension since the final surface shape is already formed very soon in the drying process. At lower temperatures, the behavior is qualitatively the same (\autoref{fig:shrinkage_30C}), where we first see lateral drying and shrinkage of the superelevated edges. The evaporation is more prevalent on the right side of the film. As with the 40~{\degree}C sample, the final film is flat with no defects at 30~{\degree}C.

In summary, the pure suspension exhibits lateral drying throughout. The edges settle faster than the center, leaving a supersaturated area that leads to drying defects. The capillary suspension shape distinctly differs during drying. Initially, the edges dry in a lateral fashion until shrinkage stops before complete edge consolidation is reached. The center nearly remains at the initial height and starts decreasing until a homogeneous surface has formed. Subsequently, the entire upper cross section homogeneously shrinks to the final coating thickness.

 \section{Discussion}

\begin{figure}[tp!]
 	\capstart
 	\centering
 	\includegraphics[width=1\linewidth]{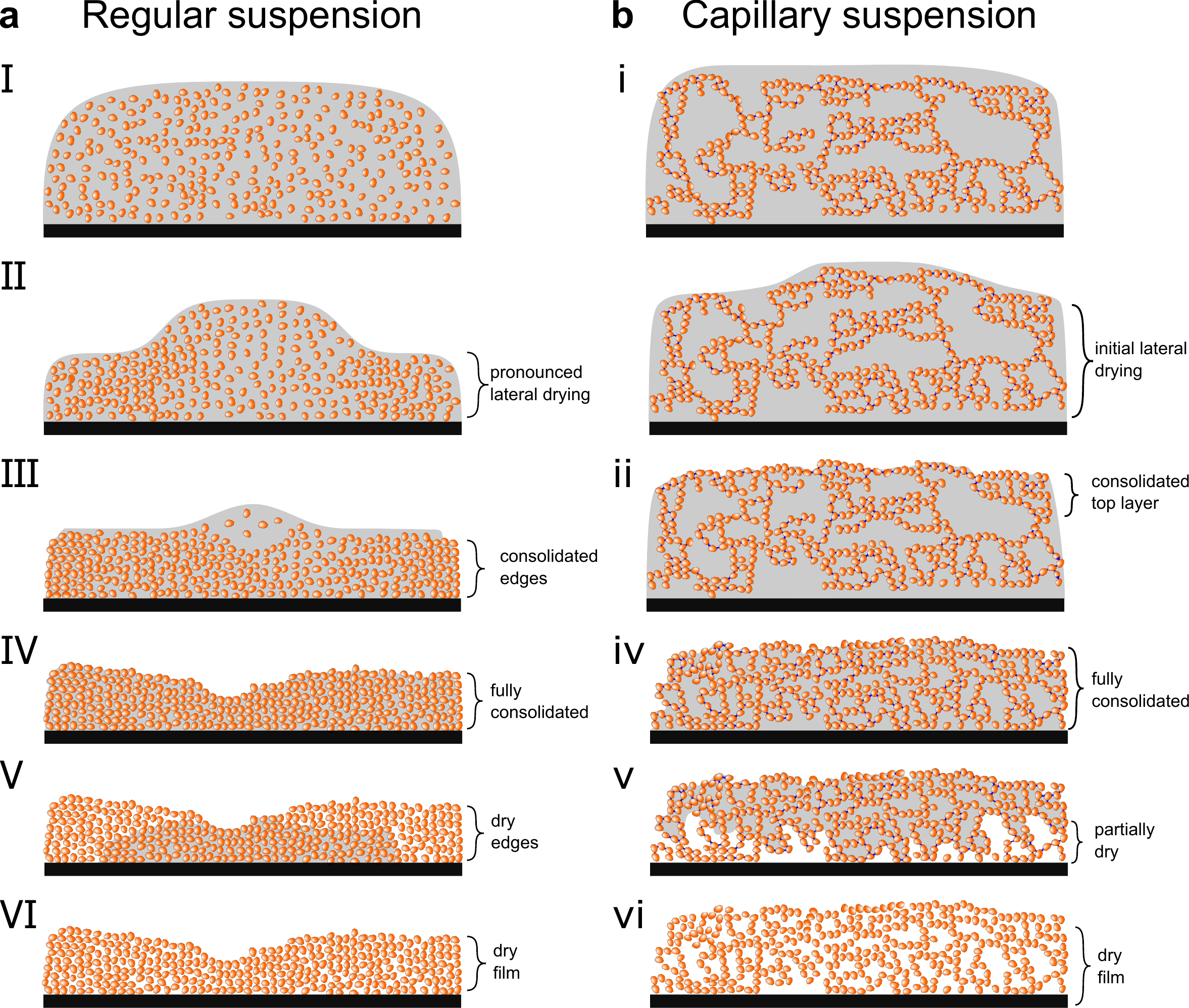}
 	\caption{ Schematic drawing of (\textbf{a}) a regular suspension without added secondary liquid and (\textbf{b}) a capillary suspension with added water. Key points in the drying process are marked using Roman numerals.}
 	\label{fig:shapesketch}
 \end{figure}

 The sketches in \autoref{fig:shapesketch}a 
 summarize the drying behavior of the regular suspension. These sketches are consistent with the experimental results and confirm the findings of other researchers for film drying of particulate coatings \cite{Bauer.2009,Chiu.1993b,Guo.1999,Jiao.2018,Tirumkudulu.2004,Holmes.2006,Routh.1998}. Initially, the film is supersaturated (\autoref{fig:shapesketch}a.I). This supersaturation is confirmed by the highly reflective surface in the undyed sample (\autoref{fig:undyed}b.I). As drying begins, the supersaturated coating transitions toward a saturated one through lateral drying (\autoref{fig:undyed}b.II). The lateral drying occurs as visually observed in the shrinking of the reflective region towards the center (\autoref{fig:undyed}b.II) and darker region in the center of the cantilever for the dyed sample in \autoref{fig:dyephotos}b.II. This is further supported by laser profiling during drying (\autoref{fig:shrinkage_40C}a). This lateral drying is uninhibited in terms of mass transport and solvent evaporates at a constant rate. As drying continues in the sedimentation regime (\autoref{fig:shapesketch}a.II), particles start settling near the edges. This denser packing is seen in \autoref{fig:undyed}a as the increase in the edge intensity beginning at II when the film enters the sedimentation regime. Increasing the particle loading at the edges also causes the fluid volume fraction to decrease, as shown in the steady decrease in bulk fluid intensity in \autoref{fig:dyephotos}a at the edges relative to the center of the cantilever. Finally, the sedimentation causes the stress to rise and the solvent evaporation to slow, as shown at point II in \autoref{fig:undyed}a and \autoref{fig:dyephotos}a. The pinned contact line causes flow towards the edges, which further increases the stress and slowing of the evaporation after point II. 
 At sufficiently high mobility, due to low particle loading and weak inter-particle interactions, flow and compressive stresses carry particles toward the edges (\autoref{fig:shapesketch}a.III), leaving a trench (\autoref{fig:shapesketch}a.IV). This particle movement, i.e stress release, is superimposed with the overall compressive stress rise, and leads to crack free films on the cantilever in this study. 
Trench formation was more pronounced at a temperature of 40~{\degree}C compared to 30~{\degree}C (\autoref{fig:shrinkage_30C}b) along with steeper superelevations. We attribute this phenomenon to the lower viscosity of the bulk phase at a raised temperature, which increases the particle mobility and a larger capillary pressure near the edges caused by deepening of pinned menisci between particles (see \autoref{eq:p_cap}) in an otherwise unchanged suspension. When drying continues beyond full compaction, air invades the pores of the film (\autoref{fig:shapesketch}a.V), decreasing the number of menisci and thus the exerted stress on the film. Once the film is virtually dry (\autoref{fig:shapesketch}a.VI), only the smallest pores in the film are saturated because of their large capillary pressure. These few, small menisci can still create extensive stresses in the film, as observed in \autoref{fig:undyed}a and \autoref{fig:dyephotos}a.
 
Capillary suspensions, on the other hand, dry very differently, as depicted in \autoref{fig:shapesketch}b. Previous studies of pendular state capillary suspensions using confocal microscopy have shown that the secondary fluid creates bridges between two particles or, if the amount of liquid is higher, funicular clusters \cite{Bossler.2016,Bossler.2017,Koos.2014,Natalia.2018}. In a consolidated coating, this corresponds to the secondary fluid being located at the throats of porous media. The connected particles form flocs, which are linked to each other, forming the backbone of a sample-spanning percolating network with a gel-like structure \cite{Bossler.2018}. These particle clusters are illustrated in \autoref{fig:shapesketch}b.i and are evident in the sample images as the initial film roughness observed in \autoref{fig:undyed}d.i. Laser profile measurements during drying of capillary suspensions (\autoref{fig:shrinkage_40C}c) revealed that lateral drying is only observed in the initial drying stage compared to the samples without added secondary fluid, as sketched in the second image in \autoref{fig:shapesketch}b. The laser profile (\autoref{fig:shrinkage_40C}c) clearly showed that lateral drying ceases in the early stages of drying, subsequently leading to a consolidated upper layer and drying top down. That is, the top layer consolidates while the particle clusters in the lower area of the cross section are still well-distributed (\autoref{fig:shapesketch}b.ii). We attribute this behavior to an increased yield stress due to increased particle-cluster interactions due to the higher particle loading where the film has consolidated. This compaction results in a locally more rigid gel network (higher compressive yield stress \cite{Brown.2002}) that counters further shrinkage of the film near the edges of the cantilever. Furthermore, the more compact upper particle layer near the edges results in smaller pores with smaller throats. These smaller throats probably act as capillaries, predominantly draining liquid from the larger pores in the supersaturated center region by capillary flow towards the edges, thus countering lateral evaporation \cite{Deegan.1997,Kharaghani.2013,Routh.2013}. 

A homogeneous consolidated layer forms at the surface across the entire cantilever, as shown in \autoref{fig:shapesketch}b.ii, when the surface supersaturation is diminished. Usually, one would expect the drying front to recede into the pores at this stage, accompanied by a decrease in drying rate caused by larger diffusional mass transport resistances \cite{Routh.1998}. However, the constant rate period for capillary suspensions persists well after point ii, actually lasting beyond the peak stress (\autoref{fig:undyed}c, \autoref{fig:dyephotos}c, and \autoref{fig:UV}a). While this would normally be associated with a liquid surface, the directly illuminated capillary suspension sample illustrated in \autoref{fig:undyed}d does not show any light reflection from point ii onward. This is quite unexpected since the regular suspension, which shows an earlier departure from the constant rate period, still shows supersaturated regions until point IV. One likely explanation is the superficially consolidated layer with menisci between the particles, which scatters the light in the capillary suspension \cite{Scherer.1990b}. This surface layer comprises a narrow pore and neck size distribution, furthering the capillary flow from the bulk to the evaporation surface. As the bulk liquid that is transported to the surface evaporates, the compressive stresses increase within the film. Once they exceed the compressive yield stress, the film starts to homogeneously shrink in height across the coating until a fully consolidated packing is reached (\autoref{fig:shapesketch}b.iv). This point should coincide with the peak stress in the measurements and a fully saturated film \cite{Guo.1999}. 

After the fully consolidated porous body has formed, the surface must be kept sufficiently wet by means of liquid transport from within the body, as inferred from the constant rate period. \citet{Scherer.1990b} theorized the existence of a funicular state in a pore, where the pore walls are covered with liquid capable of transporting fluid. Later, \citet{Laurindo.1998} describe this phenomenon in more detail. 
Our observations for a prolonged constant rate period match their theory. For preferentially wetting systems, as in this study, the critical contact angles necessary for film effects to occur can be calculated \cite{Prat.2007}. Once the coating has consolidated, the liquid-particle contact angle decreases, allowing air to invade the body until the first critical contact angle is obtained. Once the angle reaches this first critical angle, liquid fingers are formed that transport liquid to the surface, keeping the drying rate constant. The liquid flows towards the surface, generating dry pores at the substrate (\autoref{fig:shapesketch}b.v). The film stress begins to decrease in this period due to a decreasing number of filled pores exerting stress on the coating and substrate. Finally, when the second critical contact angle is reached at time v, flow ceases and the liquid starts depinning from the pore surface and air invades the pores, marking the beginning of the falling rate period \cite{Prat.2011,Yiotis.2012}. 
During this period, the film stress continues to decrease, albeit at a slightly slower rate than between point i and v, presumably due to slower drying, and thus, slower reduction of filled pores caused by vapor diffusion resistances. 

The partially dry coating bottom, formed between point iv and v in the capillary suspension, is the effect of a constant drying rate beyond the close packed coating. However, we must question why we observe transport from the substrate as well as a prolonged constant drying rate. The capillary pressure over a curved surface, described by \autoref{eq:p_cap}, is the difference between the pressure of the invading fluid (air) and the pressure of the displaced fluid (1-heptanol). For invasion to take place, a critical threshold pressure has to be exceeded \cite{Wu.2016b}.  Invasion percolation (IP) rules for hydrophilically modeled media 
\cite{Chapuis.2007, Prat.2011}, as is the case for the heptanol/water system investigated herein, dictates that invasion of a pore and throat has to be considered as two different mechanisms. Traditionally, once a throat is penetrated by air, it directly leads to the emptying of the adjacent pore and the start of porous body invasion halts the liquid flow. Thus, the obvious difference for capillary suspensions is the water bridge within the throats. These bridges must possess a higher invasion potential, and therefore, require higher air pressures before invasion occurs. Such an increase in pressure would result in a maintained corner flow and, thereby, also in a longer constant rate period for the capillary suspension. IP models for three phase flow in porous media \cite{vanDijkeM.I.J..2001}, show that during gas invasion, oil filled pores empty first by pore size in decreasing order, followed by water filled pores in the same order. In three phase flow, oil can cover water bridges that are sandwiched between the invading air \cite{Blunt.2001}. This phenomenon maintains the connection of the oil cluster, potentially resulting in better oil recovery in porous rocks. Micro-CT measurements for drainage experiments with brine, oil and gas in porous carbonate rock support the theory that invading gas predominantly drains oil, while water occupies smaller pores \cite{Scanziani.2018}. Translating these findings to drying of capillary suspensions, heptanol continues to cover the water bridges after consolidation and establishment of corner flow, leading to a connected oil cluster that persists longer than for the regular suspension, resulting in better drying. 

In the previous study by \citet{Schneider.2017} using similar samples, capillary suspensions appeared to dry significantly faster than samples without added secondary liquid. However, our current experiments seemingly contradict these observations. The discrepancy can be explained by the different methods used in each experiment. Here, we use a gravimetric technique, whereas \citet{Schneider.2017} used infrared spectroscopy (FTIR-ATR) where the sample was coated onto the crystal. The sample composition is then determined at a location no more than a few micrometers above the crystal substrate. Therefore, the previous study only measured the interface between coating and substrate. In light of the present study, we have evidence for a partially dry coating near the substrate during drying of capillary suspensions as illustrated in \autoref{fig:shapesketch}b.v, caused by corner flow and delayed air invasion into the porous body. 

A study by \citet{Jin.2013} showed very similar behavior with a significantly decreased air invasion speed as capillary suspensions. Sterically stabilized colloidal spheres were suspended in an oil phase with surfactant stabilized aqueous glycerol emulsion droplets. Using confocal microscopy, they showed that the large emulsion droplets burst during drying and the aqueous glycerol subsequently occupied the space between particles. The collapsed droplets likely occupy the throats between particles. As we described earlier, water in small pores and throats surrounded and sandwiched by oil during drying inhibit air invasion as in the drying of capillary suspensions. The result was decreased, or even absent, cracking after drying in their sample dispersion. However, the incorporation of the droplets in their suspension caused the viscous modulus $ (G'') $ to be larger than the elastic modulus $ (G') $, transitioning the formulation from a gel state towards viscous behavior. Such a viscous dominated response is undesirable for many applications such as screen printing. Upon stress build up during drying, the particles are able to release the stresses by simple migration. In lateral drying of circular droplets, this leads to a particle depleted zone or pinholes. 

While lateral drying is suppressed  and $ G'>G'' $ in the present capillary suspension system, the superelevations near the edges of the cantilever, as seen in \autoref{fig:shrinkage_40C}, remain a problem for various applications. The superelevations occur for both investigated suspensions, however they are more pronounced for capillary suspensions. These defects pose the risk of coating damage, for example during the calendaring step in the production of Li-ion batteries \cite{Schmitt.2014}. Calendaring adjusts coating porosity, and superelevations lead to a denser percolated network, negatively influencing optimal current distribution. In this study, the coating was applied on a narrow and confined cantilever, which increases edge effects from the substrate. In most applications, films are usually printed on a flat surface, such as a copper foil, avoiding the substrate edges. Previous research on printing capillary suspension slurries for Li-ion batteries showed a clear improvement of contour sharpness as well as diminished superelevations \cite{Bitsch.2014}. The otherwise very even coating dried uniformly with only little lateral drying. Such uniform drying on a cantilever has only been observed with a modified substrate. \citet{Price.2012} added walls to their cantilever, which led to a more uniform drying of the coating, forming a packed surface layer across the entire substrate similar to the current results for capillary suspensions. The walled cantilever, however, caused some particle migration, resulting in concavely shaped edges, due to the pinned contact line at the wall.

\section{Conclusions}
\label{conclusions}

On the basis of the capillary suspension concept \cite{Koos.2011, Koos.2014}, we studied the drying of alumina suspensions in 1-heptanol enhanced with a small amount of immiscible water. Compared to the regular suspension, we only observe minimal, initial lateral drying and thus no particle accumulation near the edges of a flat substrate in the capillary suspension. The initial lateral drying locally increases the yield stress, preventing particle migration and leading to a packed surface layer, while the lower regions are still suspended. Surprisingly, we found that capillary suspensions exhibit a constant drying rate period that extended beyond the fully compacted state, in contrast to typical results expected for hard sphere suspensions film drying of particulate coatings, even with binders \cite{Fu.2015}. 

The differences between the drying of the capillary suspension and pure suspension without added water is studied using simultaneous stress and weight measurements, enhanced with imaging of the cantilever during the drying process. Such simultaneous measurements, presented for the first time in this paper, allow us to monitor the drying process at higher temperatures without making assumptions about the drying conditions \cite{Payne.1997, Lei.2001, Martinez.2002b, Wedin.2004, Kim.2009}.  These measurements are combined with additional profiling of the film surface to further elucidate the changes between the formulations.  Using these combined measurements, we were able to determine that the extended constant drying rate period can be attributed to corner flow in pores near the surface. This phenomenon explains the seemingly contradictory observations of a non-reflective (i.e.~partially dry) surface with the constant drying period. The capillary bridges formed by the added water (with higher capillary pressure) prevent air invasion, extending the duration of the corner flow. Additionally, this flux leads to pore emptying near the substrate prior to the air substantially invading the top layers of a capillary suspension coating. In summary, capillary suspensions yield uniform coatings devoid of defects such as pinholes and trenches without the need for additives.

The absence of polymeric additives reduces subsequent processing steps, such as binder burn out \cite{Tsai.1991, BohnleinMau.1992}, and adds value to capillary suspensions with previously unrealized specifications on flat, narrow substrates \cite{Bauer.2009, Chiu.1993b, Guo.1999, Jiao.2018, Tirumkudulu.2004, Holmes.2006, Routh.1998}. The corner flow allows capillary suspensions to be efficiently dried under convective flow in industrial applications \cite{Cairncross.1996b}. 
	Furthermore, this capillary suspension method is capable of forming bridges with wide particle size distributions, and thus should reduce segregation of small particles during drying \cite{Bossler.2017, Dittmann.2016, Koos.2012, Maurath.2015, Schneider.2017}. Segregation between large and small particles is a particular problem for formulations with binders; different drying conditions can cause binder to settle or be transported to the surface, leading to inhomogeneous films (demixing) \cite{Jean.2001, Westphal.2015}. In capillary suspensions, the capillary bridges do not accumulate at a particular spot and small (binder) particles can even be isolated in the bridges \cite{Weiss.2020}. 

In the present work, we examined the difference between a capillary suspension and pure suspension without added water. These experiments should be expanded to include differing secondary fluid fractions to investigate the link between the particle network structuring and drying behavior in more detail, as well as extended to different drying conditions. By varying the environmental conditions as well as the vapor pressure of the secondary liquid, we could potentially change the rate at which the bridges evaporate in comparison with the bulk liquid. This could be used to potentially control both the drying rate and the final film morphology.  Finally, the present apparatus, which allows simultaneous measurements of the drying stress with the film weight and surface visualization, can be used to study other formulations. Edged cantilevers, made with photolithography, can be used to inhibit lateral drying allowing us to make a better link between the drying and formulation \cite{Price.2012}.

\section{Acknowledgements}

The authors would like to thank Almatis GmbH for the donation of alumina particles. We greatly appreciate Jana Kumberg and the entire Thin Film Technology group (TFT) of Prof. Schabel at the Karlsruhe Institute of Technology (KIT), Germany for help and access to the laser profiler. Dr. Frederik Ceyssens, from the ESAT-MICAS department at KU Leuven, Belgium is thanked for access and an introduction to the stylus surface profilometer. Finally, we acknowledge financial support from the German Research Foundation, DFG under project number KO 4805/2-1 and the Research Foundation Flanders (FWO) Odysseus Program (grant agreement no. G0H9518N).

\FloatBarrier


\label{references}
\bibliographystyle{elsarticle-num-names}
\bibliography{Suppressing}

\pagebreak


\appendix
\setcounter{figure}{0} \renewcommand{\thefigure}{S\arabic{figure}}
\setcounter{page}{1}

\begin{centering}
{\Large Supplementary Material}\\
to \\
{\Large Using an added liquid to suppress drying defects in hard particle coatings} \\
\vspace{0.5 cm}
Steffen B. Fischer$^\mathrm{a,b}$, Erin Koos$^\mathrm{a,\ast}$

\end{centering}

\noindent $^\mathrm{a}$ {\small KU Leuven, Soft Matter, Rheology and Technology, Department of Chemical Engineering, Celestijnenlaan 200f, 3001 Leuven, Belgium} \\
\noindent $^\mathrm{b}$ {\small Karlsruhe Institute of Technology, Institute for Mechanical Process Engineering and Mechanics, Karlsruhe, Germany}


\subsection*{Coating process}
\addcontentsline{toc}{section}{Coating process}

The samples were coated on stainless steel cantilevers with dimensions measuring 6~mm wide, 40~mm long, and 200~{\micro}m thick. The cantilever was clamped into a fixture between two small steel plates and placed in a milled and surface sanded coating support. The cantilever and fixture are shown in the supplementary information \autoref{fig:method}. Before coating with a ZUA 2000 doctor blade (Zehntner GmbH, Sissach, Switzerland), the samples were re-mixed for one minute at 3500 rpm in the Speedmixer, reconstituting well-mixed suspensions without agglomerates. The distinctly different rheological properties of a capillary suspension, such as shear thinning, high yield stress, as well as wall slip \cite{Koos.2014b} required varying the coating parameters for each suspension. While the best film results for the regular suspension without any added water ($\phi_\mathrm{sec}=0$), were obtained at a coating speed of 0.07~m/s and a set gap size of 260~{\micro}m. The capillary suspension ($\phi_\mathrm{sec}=0.025$) required a pre-coating step with a spatula to spread the paste. The coating speed was increased to 0.29~m/s and the coating knife gap height was reduced to 230~{\micro}m in order to obtain a similar average dry film thickness. The doctor blade was dragged over the cantilever by means of a self built film applicator driven by a voltage regulated motor. Dry film thickness was measured with a digimatic micrometer indicator (ID-H530, Mitutoyo, Japan) with a precision of 0.0015~mm. Since the dry films are very delicate, special care had to be taken in carefully lowering the measuring stylus on the coating to prevent excessive compression of the coating. The experiments reported here all had dry film thicknesses of 75~$\pm$~5~{\micro}m.

\begin{figure}[h!]
	\capstart 
	\centering
	\includegraphics[width=1\columnwidth]{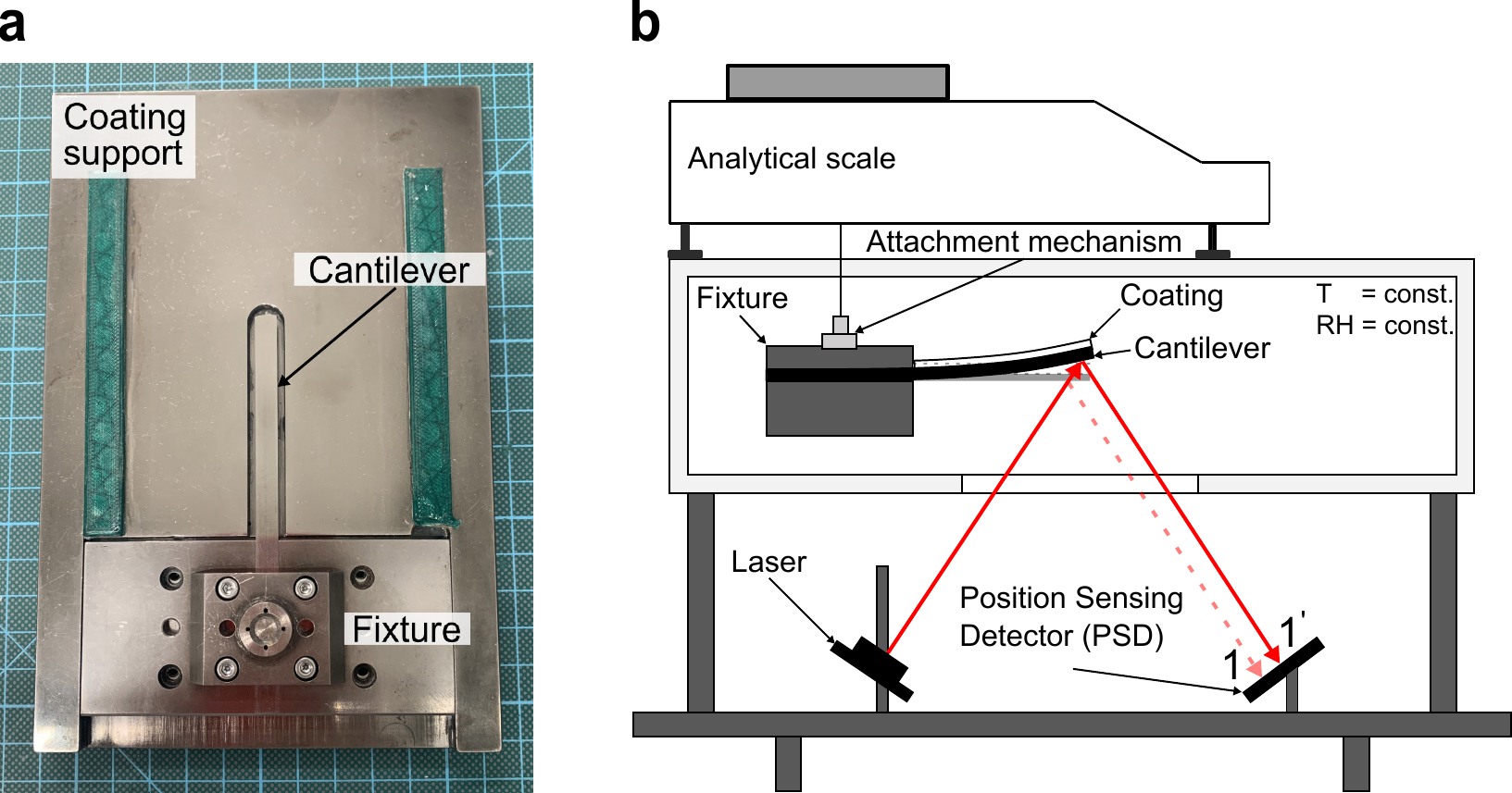}
	\caption{(\textbf{a}) Steel made coating rig with the clamped cantilever fixture placed in the recess of the rig. To maintain the cantilever’s position,	it is supported from below. Upon sample deposition, a coating blade with preset gap height is moved between the green rails along the cantilever with a constant coating velocity. (\textbf{b}) Schematic of the humidity and temperature controlled drying chamber. After coating, the fixture is inserted into the chamber, and attached to the analytical balance. Subsequently, the weight loss and cantilever deflection, tracked by the position sensing detector, are recorded.}
	\label{fig:method}
\end{figure}

\begin{figure*}[h!]
	\capstart
	\centering
	\includegraphics[width=1\linewidth]{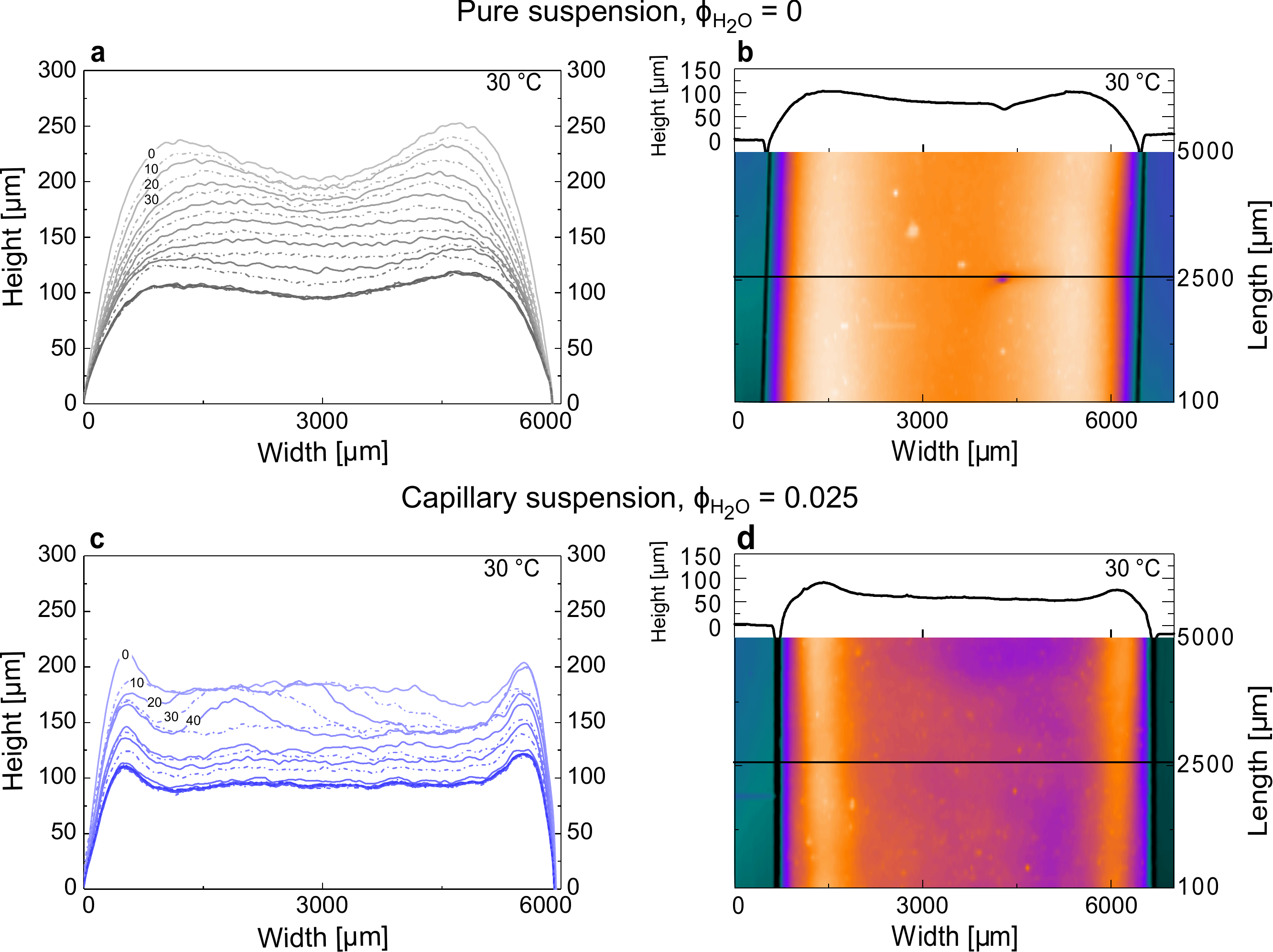}
	\caption{Film profile measurements of (\textbf{a, b}) pure suspension and (\textbf{c, d}) capillary suspension. The laser profiles during drying, shown in (\textbf{a}) and (\textbf{c}) were taken every ten minutes during contact drying at 30~{\degree}C and are displayed with solid and dashed lines in an alternating fashion. Increasing opacity of the curves indicate later stages of drying. The profiles in (\textbf{b}) and (\textbf{d}) capture a five millimeter surface section of the dried films. The horizontal black line shows the cross sectional cut for the height profile in the upper panel.}
	\label{fig:shrinkage_30C}
\end{figure*}

\begin{figure*}[h!]
	\capstart
	\centering
	\includegraphics[width=1\linewidth]{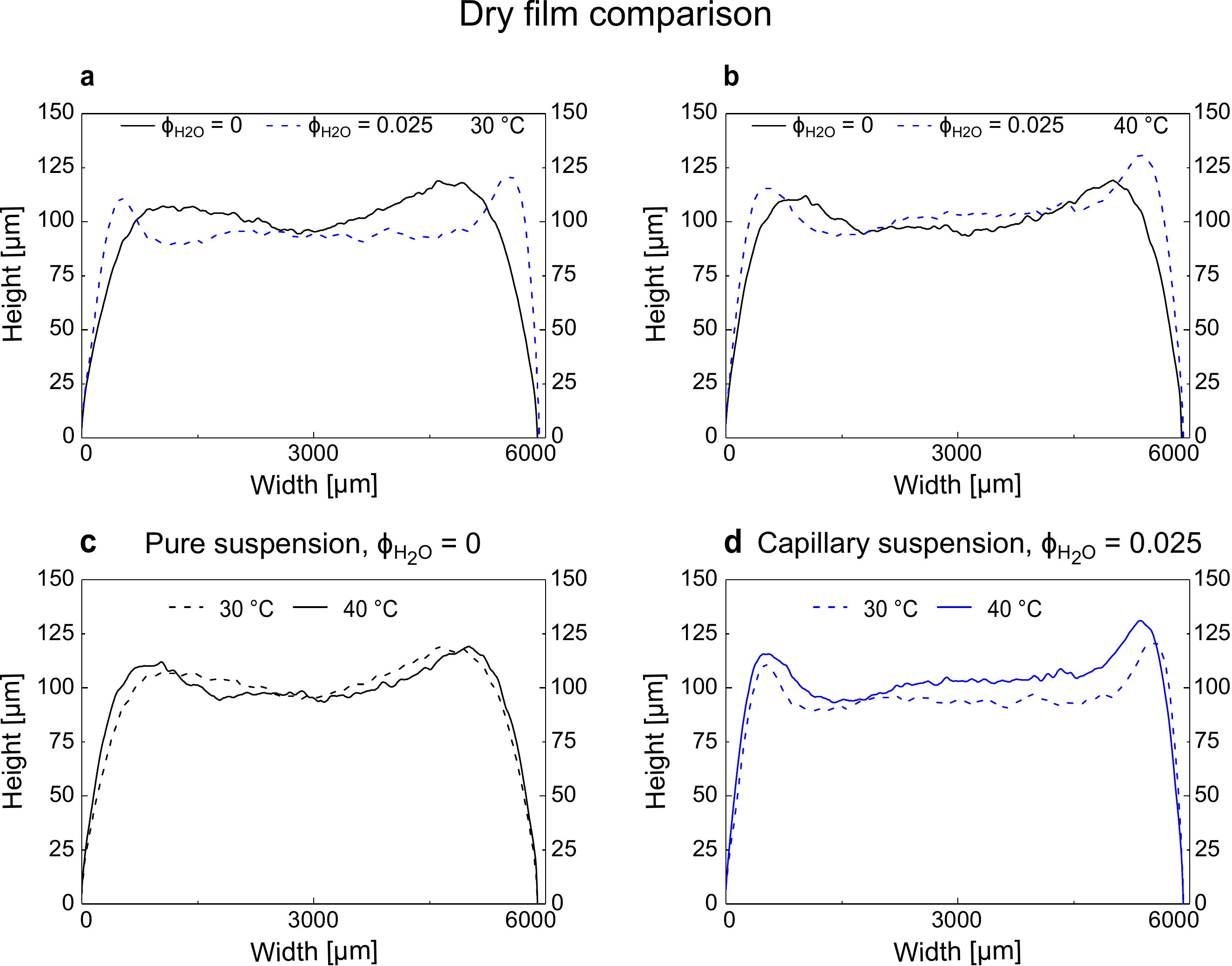}
	\caption{Dry film profilometry comparison. (\textbf{a}) Dry film of pure suspension, and regular suspension at 30~{\degree}C and (\textbf{b}) at 40~{\degree}C. (\textbf{c}) compares the dry film at 30~{\degree}C and 40~{\degree}C for the pure suspension, and (\textbf{d}) for the capillary suspension.}
	\label{fig:dry_film}
\end{figure*}









\end{document}